\documentclass{ws-ijmpa}

\begin{document}

\markboth{B. Quinn}
{Upgrade of the D$\emptyset$ Detector: The Tevatron Beyond 2 fb$^{-1}$}

%
\catchline{}{}{}{}{}
%

\title{UPGRADE OF THE D$\emptyset$ DETECTOR:\\
 THE TEVATRON BEYOND 2 fb$^{-1}$}

\author{\footnotesize BREESE QUINN\footnote{For the D$\emptyset$ Collaboration, E-mail: quinn@phy.olemiss.edu}}

\address{The University of Mississippi, University, MS 38677, USA}

\maketitle

\pub{Received (Day Month Year)}{Revised (Day Month Year)}

\begin{abstract}
Recent performance of Fermilab's Tevatron has exceeded this year's design goals and further accelerator upgrades are underway.  The high-luminosity period which follows these improvements is known as Run IIb.  The D$\emptyset$ experiment is in the midst of a comprehensive upgrade program designed to enable it to thrive with much higher data rate and occupancy.  Extensive modifications of and additions to all levels of the trigger and the silicon tracker are in progress.  All upgrade projects are on schedule for installation in the 2005 shutdown.

\end{abstract}

\section{Introduction}

Run IIa at the Tevatron will deliver $\sim$1 fb$^{-1}$ of 
integrated luminosity with peak luminosity nearly 
1.0$\times10^{32}$ cm$^{-2}$sec$^{-1}$ by mid-2005.  The D$\emptyset$ 
detector and trigger are performing very well, however aging of the inner 
silicon tracker and occupancy-related trigger rate issues will become 
areas of concern by the end of Run IIa.  The plans for Run IIb, beginning Fall 2005, are to achieve peak and integrated luminosities of 
2.8$\times10^{32}$ cm$^{-2}$sec$^{-1}$ and 8 fb$^{-1}$, respectively.\cite{lumi} 
During a Tevatron shutdown after Run IIa, the D$\emptyset$ experiment will 
complete significant detector and trigger upgrades to deal with the 
consequences of such an intense beam environment.\cite{upgrade}$^{,}$\cite{Abolins}

Of D$\emptyset$'s various subdetectors,\cite{detector} the inner silicon microstrip tracker 
(SMT) will be most severely impacted by the Run IIb integrated radiation dose.
The innermost of the four SMT silicon layers will likely die from radiation 
damage at $\sim$4 fb$^{-1}$.\cite{Lipton}  The outer central fiber tracker (CFT) will 
experience a decrease in tracking efficiency due to high track multiplicities 
corresponding to Run IIb peak luminosities.

D$\emptyset$ employs a three-level trigger architecture.  With 
a current output rate of 1.5 kHz, Level 1 (L1) consists of fast custom 
electronics making trigger decisions every 396 ns on the signal output from 
each subdetector.  The Level 2 (L2) systems apply pattern recognition 
algorithms on the data using generic processors in custom hardware, achieving 
an 1 kHz L2 accept rate.  Finally, the Level 3 (L3) Linux farm runs 
simplified reconstruction code to make selections from the L2 output for an 
event-to-tape rate of 50 Hz.  In Run IIb, D$\emptyset$ will see the average 
number of minimum bias interactions per beam crossing increase from 1 to 5, 
and the output from the current L1 system would reach 30 kHz.  All three 
trigger levels will need to be enhanced in order to handle the high 
occupancies and rates of Run IIb. 

\section{Improved Detector Performance}

To compensate for the loss of silicon devices in the SMT from radiation damage and other aging effects, a new radiation-hard inner silicon Layer 0 (L0) will be installed on the beam pipe at R = 1.6 cm.\cite{L0design}  The additional L0 coverage will help to recover losses in tracking and b-tagging efficiency that result from dead regions, particularly in the inner layer of the Run IIa SMT.  The improvement in impact parameter resolution with L0 compared to the Run IIa SMT with and without its inner layer is shown in Fig. 1.  The smaller resolution that results from including L0 hits near the interaction point corresponds to a 15\% improvement over Run IIa b-tagging efficiency.  The L0 design utilizes the new SVX4 readout chip and a novel grounding scheme which provides excellent noise reduction by covering all carbon fiber support surfaces with a laminated ground mesh.\cite{Garcia}$^{,}$\cite{Quinn} 

\begin{figure}
\centerline{\psfig{file=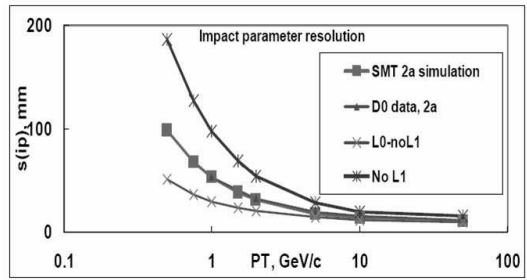,width=7cm}}
\vspace*{8pt}
\caption{Impact parameter resolution for the Run IIa SMT with and without the inner layer (L1), and with the inclusion of L0.}
\end{figure}

\section{Increased Trigger Rejection at L1}

The most ambitious trigger system upgrade is the replacement of the entire L1 calorimeter trigger (L1CAL).  The new L1CAL uses digital filtering to improve resolution on the measurement of transverse energy and a sliding window algorithm to perform better clustering.\cite{Bystricky}  These techniques will dramatically sharpen the trigger threshold turn-on curves and bring much of the current L2 rejection up to L1. 

The current central track trigger (L1CTT) employs fiber doublet logic to find CFT track candidates which are then passed on to the muon and silicon trigger systems.  The upgraded L1CTT will use fiber singlets to reduce track candidate combinatorics and improve fake track rejection at high occupancies.  This will be accomplished with new digital front end boards that include larger FPGA's capable of handling more complex singlet logic algorithms.

D$\emptyset$ will add a completely new trigger system to match calorimeter clusters to CFT tracks (L1CALTRK).  This system will enhance triggers for Higgs searches by improving fake electron rejection by a factor of two and fake tau rejection by ten.  It is a straightforward project which only involves modest modifications to the existing L1 muon-CFT track matching system.

\section{More Processing Power and Reduced Backgrounds at L2 and L3}

The D$\emptyset$ trigger includes a L2 silicon track trigger (L2STT) that finds tracks in the SMT.  The L2STT will be expanded with additional silicon trigger boards identical to existing ones.  The additional boards will process the inputs from the new L0 silicon channels.  This extra information enhances pattern recognition crucial to the separation of background from signal topologies.

The L2 systems were designed to use Alpha-based single board computers for pattern recognition.  These were recently replaced with more powerful L2$\beta$ processors, and for Run IIb D$\emptyset$ will install additional new L2$\beta$'s that can employ more complex algorithms for improved background rejection.

Several upgrades to the data acquisition and online computing systems will increase D$\emptyset$'s capacity to record more high-quality data.  The most significant of these projects will be the addition of 96 Linux nodes to the L3 computing farm.  The expansion will effectively double the L3 processing power which will confer the ability to efficiently process the more complex high luminosity Run IIb events.  In addition,the L2 accept rate will be raised to 1 kHz.

\section{Status and Conclusions}

The D$\emptyset$ upgrades for Run IIb are maintaining a schedule for installation during the Tevatron shutdown in 2005.  All electronics for the trigger upgrades are either in final prototype testing or ordered for production, and the L0 project is ahead of schedule for early production of silicon modules.  Significant resources are now being brought to bear on the software and commissioning efforts required to ensure a quick return to physics quality data taking when Run IIb begins.

\end{document}